\documentclass[conference]{IEEEtran}
\usepackage{cite}

\usepackage{graphicx}
\usepackage{fancyhdr}
\usepackage[misc]{ifsym}
\usepackage{amsmath,amssymb,amstext}
\usepackage{bm}
\usepackage{stfloats}
\usepackage{algorithm}
\usepackage{multicol}
\usepackage{subfigure}

\usepackage{tabulary}
\usepackage{booktabs}

\newcommand{\rank}{\ensuremath{\operatorname{rank}}}

\ifCLASSINFOpdf
\else
\fi
\begin{document}
%
\title{Secrecy Beamforming for SWIPT MISO Heterogeneous Cellular Networks
}



\author{\IEEEauthorblockN{Hui Ma\IEEEauthorrefmark{1}, Julian Cheng\IEEEauthorrefmark{1} and Xianfu Wang\IEEEauthorrefmark{2}}
\IEEEauthorblockA{\IEEEauthorrefmark{1}School of Engineering, The University of British Columbia, Kelowna, BC, Canada\\
\IEEEauthorrefmark{2} Department of Mathematics, Irving K. Barber School of Arts and Sciences,\\
The University of British Columbia, Kelowna, BC, Canada\\
Email: {hui.ma@alumni.ubc.ca}, {julian.cheng@ubc.ca}, {shawn.wang@ubc.ca}}}

\maketitle

\begin{abstract}
In this paper, we consider the secure transmission design for a multiple-input single-output Femtocell overlaid with a Macrocell in co-channel deployment. The Femtocell base station sends confidential messages to information receiving Femtocell users (FUs) and energy signals to energy receiving (ER) FUs while limiting the interference to Macrocell users (MUs). The ER FUs have the potential to wiretap the confidential messages. By taking fairness into account, we propose a sum logarithmic secrecy rate maximization beamforming design problem under the interference constraints for MUs and energy harvesting (EH) constraints for ER FUs. The formulated design problem is nontrivial to solve due to the nonconvexity which lies in the objective and the constraints. To tackle the design problem, a semidefinite relaxation and successive convex approximation based algorithm is proposed. Simulation results demonstrate the effectiveness of the proposed beamforming design.
\end{abstract}


%
\IEEEpeerreviewmaketitle

\section{Introduction}
\label{sect:introduction}

Compared with the conventional homogeneous
networks, heterogeneous cellular networks (HCNs), in which the small cells
are deployed over the macrocell coverage area, can provide better coverage and higher throughput.
Therefore, HCNs have been
regarded as a key network architecture for the fifth generation (5G) wireless communication systems \cite{Sheng2}. 
In the last a few decades, physical-layer security (PLS), which aims at exploiting the physical characteristics of wireless channels to defend against wiretapping \cite{Ren11}, has emerged as a technique for secure information transmission. Consequently, PLS for HCNs has attracted much research
attention \cite{Ren19}.

On the other hand, simultaneous wireless information and power
transfer (SWIPT), which enables mobile devices to harvest energy from ambient radio frequency (RF) signals, can provide cost-effective and perpetual power supplies for mobile devices. Therefore, SWIPT has been considered as a promising approach to the address the energy scarcity issue \cite{CZ12}. Most recently, secure transmission design with SWIPT was considered for HCNs \cite{Ren,Sys}. 
However, to the authors' best knowledge, existing works on secure transmission design for SWIPT enabled HCNs often assumed that there is only one information receiver (IRer) that should be protected from eavesdropping, which motivates us to consider the scenario where there are multiple IRers that need the protection against eavesdropping.

In this paper, we consider a Femtocell overlaid with a Macrocell in co-channel deployment. The Femtocell base station (FBS) is equipped with multiple antennas while Femtocell users (FUs) and Macrocell users (MUs) are equipped with single antenna. To mitigate the interference to MUs, we assume that the FBS has the cognitive radio capability \cite{Ren10}, where FBS can transmit simultaneously with MBS when its interference to MUs is strictly less than a predefined threshold. On the other hand, there exist two types of FUs, namely, information receiving (IR) FUs and energy receiving (ER) FUs. The FBS transmits different confidential messages to IR FUs and energy signals to ER FUs. The IR FUs that have access to information services with authorization are trust worthy, whereas the confidential messages are at the risk of being eavesdropped by the ER FUs. Note that the energy signals sent to ER FUs can also act as artificial noise (AN) that can cripple ER FUs' interception capabilities. In order to ensure fairness between IR FUs, proportional fairness is introduced. Consequently, we propose a sum logarithmic
secrecy rates maximization beamforming design problem under the interference constraints for MUs and energy harvesting constraints for ER FUs.
To deal the beamforming design problem which is nonconvex, we propose an algorithm based on the semidefinite relaxation (SDR) and successive convex approximation (SCA) techniques. The convergence of the proposed algorithm and the effectiveness of the proposed beamforming design are illustrated in simulation results.

{\emph{Notations}}-Vectors are denoted by boldface lowercase letters while matrices are denoted by boldface uppercase letters.  ${({\cdot})^T}$ represents the transpose; ${({\cdot})^H}$ represents the conjugate transpose; $\left| \cdot \right|$ represents the modulus of a complex number;  $\rm{E}[\cdot]$ represents the expectation and ${{\rm{tr}}( \cdot )}$  denotes the trace operator. The notation ${{\bf{A}} \succeq {\bf{B}}}$ implies ${{\bf{A}} - {\bf{B}}}$ is positive semidefinite. ${{\bf{0}}}$ represents a null matrix with suitable dimension. 

\section{System Model}
\label{sect:system:model}

We consider a multiuser MISO downlink femtocell overlaid with a macrocell in co-channel deployment where there exist $M$ FUs in femtocell and $N$ MUs in microcell. We assume that FBS has $T$ antennas with $T > M$, while FUs and MUs are equipped with single antenna. Furthermore, there exist two types of FUs, namely, IR FUs
that receive information from FBS and ER FUs that harvest energy from FBS. The IR FUs that
have access to information services with authorization are
trust worthy, whereas the ER FUs may accidentally eavesdrop the messages for IR FUs\footnote{The case where an IR FU may also eavesdrop on the information for the other IR FUs can be dealt with by slightly modifications of the proposed system model and algorithm.}. IR FUs, ER FUs and MUs are denoted by the sets ${\mathcal{J}} = \{1, \ldots, J\}$, ${\mathcal{K}} = \{1,\ldots, K \}$ and ${\mathcal{N}} = \{1, \ldots, N\}$ respectively. We assume single stream beamforming at FBS for information transmission. In addition, without loss of generality, we assume $K$ ER FUs are assigned with $\gamma$ energy beams ($\gamma \le T$). Therefore, the transmitted signals from FBS can be
expressed as
\begin{equation}
\label{TransSig}
{\bf{x}} = \sum\limits_{j \in {\mathcal{J}}} {{{\bf{w}}_j}s_j^{IR}}  + \sum\limits^{\gamma}_{i = 1} {{{\bf{q}}_i}s_i^{ER}}
\end{equation}
where ${{\bf{w}}_j} \in {\mathbb{C}}^{T \times 1}$ and ${{\bf{q}}_i} \in {\mathbb{C}}^{T \times 1}$ represent the information beamforming vector and the $i$-th energy beamforming vector, respectively; $s^{IR}_j$ denotes the information-bearing signal intended for the  $j$-th IR FU, while $s^{ER}_i$ denotes the $i$-th energy-carrying signal. It is assumed that $s^{IR}_j$'s are independent and identically distributed (i.i.d.) circularly symmetric
complex Gaussian (CSCG) random variables with zero mean and unit variance, i.e., $s^{IR}_j \sim \mathcal{CN}(0,1)$. Furthermore, $s^{ER}_i$'s can be arbitrary independent random signals with unit average power. Since in this paper we consider secret information transmission to the IR FUs, the energy signals $s^{ER}_i$'s also play the role of AN to reduce the information rate eavesdropped by ER FUs. 
We assume that $s^{ER}_i$'s are i.i.d. CSCG random variables denoted by $s^{ER}_i \sim \mathcal{CN}(0,1)$, since the worst-case noise distribution for the eavesdropping ER FUs is known to be Gaussian. Let $P_{max}$ denote
the maximal transmit power for FBS. Thus, based on \eqref{TransSig} we know $E[{{\bf{x}}^H}{\bf{x}}] = \sum\nolimits_{j \in {\mathcal{J}}} {{{\left\| {{{\bf{w}}_j}} \right\|}^2}}  + {\sum\nolimits^{\gamma}_{i = 1} {\left\| {{{\bf{q}}_i}} \right\|} ^2} \le P_{max}$.

We assume a quasi-static fading environment. Denote ${\bf{h}}_j \in {\mathbb{C}}^{T \times 1}$ and ${\bf{g}}_k \in {\mathbb{C}}^{T \times 1}$  as the channel vectors from FBS to the $j$-th IR FU and the $k$-th ER FU respectively, where ${\left\| {{{\bf{h}}_j}} \right\|^2} = {\rho _{h,j}}$ and ${\left\| {{{\bf{g}}_k}} \right\|^2} = {\rho _{g,k}}$ with ${\rho _{g,k}}>{\rho _{h,j}}$. In other words, we assume that ER FUs are located nearer to FBS than IR
FUs due to a higher received power requirement of energy harvesting
for real-time operation. 
Here, we assume ER FUs are active devices, which can communicate with FBS in the
uplink and harvest energy from FBS in the downlink. When FBS is a time division duplexing (TDD) system, it can estimate
the channel state information (CSI) in the uplink transmission and then acquire the CSI from FBS to ER FUs via channel
reciprocity. As for a frequency division duplexing (FDD) system, ER FUs can estimate the CSI from FBS to ER FUs in
the downlink transmission and feedback this CSI to FBS in the uplink transmission. Then, the received discrete-time baseband signals at the $j$-th IR FU and the $k$-th ER FU can be given by, respectively,
\begin{equation}
\label{IRreceive}
y_{IR,j} = {\bf{h}}_j^H{\bf{x}} + I_{IR,j} + n_{IR,j}, \forall j \in {\mathcal{J}},
\end{equation}
\begin{equation}
\label{ERreceive}
y_{ER,k} = {\bf{g}}_k^H{\bf{x}} + I_{ER,k} + n_{ER,k}, \forall k \in {\mathcal{K}}
\end{equation}
where $n_{IR,j} \sim \mathcal{CN}(0, \sigma^2_{IR,j})$ and $n_{ER,k} \sim \mathcal{CN}(0, \sigma^2_{ER,k})$ are the i.i.d. Gaussian noise terms at the  $j$-th IR FU and the  $k$-th ER FU respectively. Besides, $I_{IR,j}$
and $I_{ER,k}$ represent the interference generated by MBS at the $j$-th IR FU and the $k$-th ER FU respectively.

According to \eqref{IRreceive}, the
signal-to-interference-plus-noise ratio (SINR) at the $j$-th IR
FU can be expressed as
\begin{equation}
\label{SINRIR}
SINR_{IR,j} = \frac{{{{\left| {{\bf{h}}_j^H{{\bf{w}}_j}} \right|}^2}}}{{\sum\limits_{l \in {\mathcal{J}}, l \ne j} {{{\left| {{\bf{h}}_j^H{{\bf{w}}_l}} \right|}^2}}  + {\bf{h}}_j^H{\bf{Q}}{{\bf{h}}_j} + P_{IR,j} + \sigma _{IR,j}^2}}
\end{equation}
where ${\bf{Q}} = \sum\nolimits_{i = 1}^{\gamma} {{{\bf{q}}_i}{\bf{q}}_i^H}$ and $P_{IR,j}$ is the power of the interference caused by MBS at the $j$-th IR FU.

From \eqref{ERreceive}, the SINR at the $k$-th ER FU (suppose that it is an eavesdropper who intends
to decode the message for the $j$-th IR FU instead of harvesting
energy) can be expressed as
\begin{equation}
\label{SINRER}
SIN{R_{ER,k,j}} = \frac{{{{\left| {{\bf{g}}_k^H{{\bf{w}}_j}} \right|}^2}}}{{{{\sum\limits_{l \in {\mathcal{J}}, l \ne j} {\left| {{\bf{g}}_k^H{{\bf{w}}_l}} \right|} }^2} + {\bf{g}}_k^H{\bf{Qg}}_k + {P_{ER,k}} + \sigma _{ER,k}^2}}
\end{equation}
where $P_{ER,k}$ is the power of the interference caused by MBS at the $k$-th ER FU.

The achievable secrecy rate at the $j$-th IR FU is thus given by 
\begin{equation}
\label{Secrecyrate}
{R_j} = \log_2 \left( {1 + SIN{R_{IR,j}}} \right) - \mathop {\max }\limits_{k \in {\mathcal{K}} } \log_2 (1 + SIN{R_{EV,k,j}}).
\end{equation}

On the other hand, for wireless energy transfer, owing to the broadcast property
of wireless channels, the energy carried by information
and energy beams and the interference from MBS 
can be harvested by each ER
FU. Therefore, the harvested power $E_k$ at the $k$-th ER FU is
proportional to the total received power, and it can be written as
\begin{equation}
\label{Harvest}
E_{k} = {\xi _k}\left( {\sum\limits_{j \in {\mathcal{K}}} {{{\left| {{\bf{g}}_k^H{{\bf{w}}_j}} \right|}^2}}  + {\bf{g}}_k^H{\bf{Qg}}_k + {P_{ER,k}}} \right)
\end{equation}
where ${\xi _k}$ represents the energy harvesting efficiency and ${P_{ER,k}}$ denotes the interference from MBS to the $j$-th ER FU.

Furthermore, to mitigate the cross-tier interference to MUs in Macrocell, FBS coexists with
MUs via the underlay cognitive radio paradigm \cite{Ren10},
where FBS can transmit data with MBS simultaneously provided that the interference incurred by FBS to MUs is less than a
predefined threshold. The interference to MUs can be expressed as
\begin{equation}
\label{MUinf}
{I_{MU,n}} = \sum\limits_{j \in \mathcal{J}} {{{\left| {{\bf{i}}_n^H{{\bf{w}}_j}} \right|}^2}}  + {\bf{i}}_n^H{\bf{Qi}}_n
\end{equation}
where ${\bf{i}}_n \in  {\mathbb{C}}^{T \times 1}$ denotes the channel vector between the FBS and the $n$-th MU.

\section{Beamforming Design}
\label{sect:problem:formulation}
By considering secrecy rate fairness among users, in this
paper, we maximize the summation of the logarithmic secrecy rates of users. 
The proportional fairness based secrecy beamforming design problem is formulated as follows
\begin{subequations}
\label{mainpro}
\begin{align}
&\mathop {\max }\limits_{{{\bf{w}}_j},\forall j \in \mathcal{J},{{\bf{Q}}}} f({{\bf{w}}_j},\forall j \in \mathcal{J},{{\bf{Q}}})=\sum\limits_{j \in \mathcal{J}} {\ln ({R_j})} \\
& {\text{s.t.}} \, {E_k}  \ge {\varpi _k}, \, \forall k \in {\mathcal{K}}, \label{mainpro_C1}\\
&\hspace{0.5cm}{I_{MU,n}}  \le {\eta _n}, \, \forall n \in {\mathcal{N}}, \label{mainpro_C2}\\
&\hspace{0.5cm}\sum\limits_{j \in {\mathcal{J}}} {{{\left\| {{{\bf{w}}_j}} \right\|}^2}}  + {\rm{tr}}({\bf{Q}}) \le {P_{\max }}, \label{mainpro_C3}\\
&\hspace{0.5cm}{\bf{Q}} \succeq {\bf{0}} \label{mainpro_C4}
\end{align}
\end{subequations}
where ${\varpi _k}$ and ${\eta_n}$ denote the energy harvesting threshold for the $k$-th ER FU and the permissible interference
threshold for the $n$-th MU respectively.
Note that we optimize ${\bf{Q}}$ instead of ${\bf{q}}_i$ in \eqref{mainpro}. However, from the solution of ${\bf{Q}}$, the number of energy beams $\gamma$ can be derived as $\gamma=\rank ({{\bf{Q}}})$ and the
energy beams ${\bf{q}}_i$ can be obtained by the eigenvalue decomposition (EVD) of ${\bf{Q}}$.

It is nontrivial to solve the beamforming design problem \eqref{mainpro} owing to the non-convexity lying in the objective
function and some constraints. Hence, we deal with problem \eqref{mainpro} by employing the SDR and SCA techniques.
To make the original design problem \eqref{mainpro} more tractable, we define ${{{\bf{H}}_j}}={{\bf{h}}_j}{{\bf{h}}_j^H}$, ${{{\bf{G}}_k}}={{\bf{g}}_k}{{\bf{g}}_k^H}$, ${{{\bf{I}}_n}}={{\bf{i}}_n}{{\bf{i}}_n^H}$ and ${{{\bf{W}}_j}}={{\bf{w}}_j}{{\bf{w}}_j^H}$. 
Then, it follows that $\rank({{{\bf{W}}_j}}) = 1, \, \forall j \in \mathcal{J}$.  
By ignoring the rank one constraints on ${{{\bf{W}}_j}}$'s, the SDR problem of \eqref{mainpro} can be expressed as
\begin{subequations}
\label{SDR}
\begin{align}
&\mathop {\max}\limits_{{{\bf{W}}_j},\forall j \in \mathcal{J},{\bf{Q}}} \, {\bar{f}}({{{\bf{W}}_j},\forall j \in \mathcal{J},{\bf{Q}}}) \nonumber\\
&\hspace{0cm}=\sum\limits_{j\in \mathcal{J}} \ln {{\Big{(}} {\log_2 \left( {1 + \overline{SINR}_{IR,j}} \right)} } \nonumber\\
&\hspace{2.5cm} { - \mathop {\max }\limits_{k\in \mathcal{K}} \log_2 \left( {1 + \overline{SINR}_{ER,k,j}} \right)} {\Big{)}} \label{SDR_Ob}\\
&{\text{s.t.}} \, {\xi _k}\left( {\sum\limits_{j \in \mathcal{J}} {tr\left( {{\bf{G}}_k^{}{{\bf{W}}_j}} \right)}  + tr\left( {{\bf{G}}_k^{}{\bf{Q}}} \right) + {P_{ER,k}}} \right) \ge {\varpi _k}, \, \forall k\in \mathcal{K}, \label{SDR_C1}\\
&\hspace{0.5cm} \, \sum\limits_{j\in \mathcal{J}} {tr\left( {{\bf{I}}_n^{}{{\bf{W}}_j}} \right)}  + tr\left( {{\bf{I}}_n^{}{\bf{Q}}} \right) \le {\eta _n}, \, \forall n\in \mathcal{N}, \label{SDR_C2}\\
&\hspace{0.5cm} \,  \sum\limits_{j\in \mathcal{J}} {tr\left( {{{\bf{W}}_j}} \right)}  + {\rm{tr}}({\bf{Q}}) \le {P_{\max }}, \label{SDR_C3}\\
&\hspace{0.5cm} \, {\bf{Q}} \succeq {\bf{0}}, {{\bf{W}}_j} \succeq {\bf{0}}, \, \forall j\in \mathcal{J} \label{SDR_C4},
\end{align}
\end{subequations}
where
\begin{equation}
\begin{split}
\nonumber
&\overline{SINR}_{IR,j} = \\
&\hspace{1.2cm}\frac{{tr({\bf{H}}_j^{}{{\bf{W}}_j})}}{{\sum\limits_{l\in \mathcal{J}, l\ne j} {tr\left( {{\bf{H}}_j^{}{{\bf{W}}_l}} \right)}  + tr\left( {{\bf{H}}_j^{}{\bf{Q}}} \right) + {P_{IR,j}} + \sigma _{IR,j}^2}},
\end{split}
\end{equation}
\begin{equation}
\begin{split}
\nonumber
&\overline{SINR}_{ER,k,j} = \\
&\hspace{1.2cm}\frac{{tr\left( {{\bf{G}}_k^{}{{\bf{W}}_j}} \right)}}{{\sum\limits_{l\in \mathcal{J}, l \ne j} {tr\left( {{\bf{G}}_k^{}{{\bf{W}}_l}} \right)}  + tr\left( {{\bf{G}}_k^{}{\bf{Q}}} \right) + {P_{ER,k}} + \sigma _{ER,k}^2}}.
\end{split}
\end{equation}

The SDR problem \eqref{SDR}, however, is still challenging to solve, since the
objective function \eqref{SDR_Ob} is nonconvex and has complicated form. Therefore, we deal with  problem \eqref{SDR} by employing the SCA approach.

First, let us introduce some auxiliary variables as follows
\begin{subequations}
\label{RelVar}
\begin{align}
&{2^{{a_j}}}=\sum\limits_{l \in \mathcal{J}} {tr\left( {{\bf{H}}_j^{}{{\bf{W}}_l}} \right)}  + tr\left( {{\bf{H}}_j^{}{\bf{Q}}} \right) + {P_{IR,j}} + \sigma _{IR,j}^2, \, \forall j \in \mathcal{J}, \label{RelVara}\\
&{2^{{b_j}}}=\sum\limits_{l \in \mathcal{J}, l \ne j} {tr\left( {{\bf{H}}_j^{}{{\bf{W}}_l}} \right)}  + tr\left( {{\bf{H}}_j^{}{\bf{Q}}} \right) + {P_{IR,j}} + \sigma _{IR,j}^2, \, \forall j \in \mathcal{J}, \label{RelVarb}\\
&{2^{{c_{k}}}} = \sum\limits_{l \in \mathcal{J}} {tr\left( {{\bf{G}}_k^{}{{\bf{W}}_l}} \right)}  + tr\left( {{\bf{G}}_k^{}{\bf{Q}}} \right) + {P_{ER,k}} + \sigma _{ER,k}^2, \forall k \in \mathcal{K},  \label{RelVarc}\\
&2^{{d_{k,j}}}=\sum\limits_{l \in \mathcal{J}, l \ne j} {tr\left( {{\bf{G}}_k^{}{{\bf{W}}_l}} \right)}  + tr\left( {{\bf{G}}_k^{}{\bf{Q}}} \right) + {P_{ER,k}} + \sigma _{ER,k}^2, \nonumber\\
&\hspace{5.2cm} \forall k \in \mathcal{K}, \forall j \in \mathcal{J}. \label{RelVard}
\end{align}
\end{subequations}
From \eqref{RelVara}-\eqref{RelVarb}, we rewrite \eqref{SDR} as
\begin{equation}
\label{SDR_middle}
\begin{split}
&\mathop {\max}\limits_{{{\bf{W}}_j},{\bf{Q}},r_j,a_j,b_j,c_{k},d_{k,j},\forall k \in \mathcal{K},\forall j \in \mathcal{J}} \sum\limits_{j\in \mathcal{J}}{\ln (r_j)} \\
&{\text{s.t.}} \, {a_j} - {b_j} - {c_{k}} + {d_{k,j}} \ge {r_j}, \forall k \in \mathcal{K}, \forall j \in \mathcal{J},  \\
&\hspace{0.5cm}\eqref{RelVara}- \eqref{RelVard}, \\
&\hspace{0.5cm} \eqref{SDR_C1}-\eqref{SDR_C4}
\end{split}
\end{equation}
By relaxing the equality constraints \eqref{RelVara}- \eqref{RelVard}, we can transform \eqref{SDR_middle} into
\begin{subequations}
\label{SDR_relax}
\begin{align}
&\mathop {\max }\limits_{{{\bf{W}}_j},{\bf{Q}},r_j,a_j,b_j,c_{k},d_{k,j},\forall k \in \mathcal{K}, \forall j \in \mathcal{J}} \sum\limits_{j\in \mathcal{J}}{\ln (r_j)} \\
&{\text{s.t.}} \, {a_j} - {b_j} - {c_{k}} + {d_{k,j}} \ge {r_j}, \label{SDRR_C1}\\
&\hspace{0.5cm} \sum\limits_{l \in \mathcal{J}} {tr\left( {{\bf{H}}_j^{}{{\bf{W}}_l}} \right)}  + tr\left( {{\bf{H}}_j^{}{\bf{Q}}} \right) + {P_{IR,j}} + \sigma _{IR,j}^2\ge {2^{{a_j}}}, \nonumber\\
&\hspace{6.5cm} \forall j \in \mathcal{J}, \label{SDRR_C2}\\
&\hspace{0.5cm} \sum\limits_{l \in \mathcal{J}, l \ne j} {tr\left( {{\bf{H}}_j^{}{{\bf{W}}_l}} \right)}  + tr\left( {{\bf{H}}_j^{}{\bf{Q}}} \right) + {P_{IR,j}} + \sigma _{IR,j}^2 \le {2^{{b_j}}} ,\nonumber\\
&\hspace{6.5cm} \forall j \in \mathcal{J}, \label{SDRR_C3}\\
&\hspace{0.5cm} \sum\limits_{l \in \mathcal{J}} {tr\left( {{\bf{G}}_k^{}{{\bf{W}}_l}} \right)}  + tr\left( {{\bf{G}}_k^{}{\bf{Q}}} \right) + {P_{ER,k}} + \sigma _{ER,k}^2 \le {2^{{c_{k}}}}, \nonumber\\
&\hspace{6.5cm} \forall k \in \mathcal{K}, \label{SDRR_C4}\\
&\hspace{0.5cm} \sum\limits_{l \in \mathcal{J}, l \ne j} {tr\left( {{\bf{G}}_k^{}{{\bf{W}}_l}} \right)}  + tr\left( {{\bf{G}}_k^{}{\bf{Q}}} \right) + {P_{ER,k}} + \sigma _{ER,k}^2 \ge {2^{{d_{k,j}}}}, \nonumber\\
&\hspace{5.2cm} \forall j \in \mathcal{J}, \forall k \in \mathcal{K}, \label{SDRR_C5}\\
&\hspace{0.5cm} \eqref{SDR_C1}-\eqref{SDR_C4}. \nonumber
\end{align}
\end{subequations}
It can be observed that \eqref{SDRR_C2}-\eqref{SDRR_C5} are always active constraints. Therefore, eq. \eqref{SDR_relax} is equivalent to \eqref{SDR_middle}.

Constraints \eqref{SDRR_C3} and \eqref{SDRR_C4} result in the non-convexity of \eqref{SDR_relax}. Therefore, we need to convexify these constraints via approximation techniques. 
Suppose ${{\bar{b}}_j}$ and ${{\bar{c}}_{k}}$ are feasible to \eqref{SDR_relax}. Since ${2^{{b_j}}}$ and ${2^{{c_{k}}}}$ are both convex, according to the first-order Taylor expansion, we have ${2^{{b_j}}} \ge  {2^{{\bar{b}}_j}}(\ln(2){b_j} -
\ln(2){\bar{b}}_j + 1)$ and  ${2^{{c_{k}}}} \ge {2^{{\bar{c}}_{k}}}(\ln(2){c_{k}} -
\ln(2){\bar{c}}_{k} + 1)$. Thus, constraints \eqref{SDRR_C3} and \eqref{SDRR_C4} can be conservatively approximated at  ${{\bar{b}}_j}$ and ${{\bar{c}}_{k}}$ as
\begin{equation}
\begin{split}
\label{SDRR_C3_Appro}
&\sum\limits_{l \in \mathcal{J}, l \ne j} {tr\left( {{\bf{H}}_j^{}{{\bf{W}}_l}} \right)}  + tr\left( {{\bf{H}}_j^{}{\bf{Q}}} \right) + {P_{IR,j}} + \sigma _{IR,j}^2  \le \\
&\hspace{4cm}{2^{{\bar{b}}_j}}(\ln(2){b_j} - \ln(2){\bar{b}}_j + 1)
\end{split}
\end{equation}
and
\begin{equation}
\begin{split}
\label{SDRR_C4_Appro}
&\sum\limits_{l \in \mathcal{J}} {tr\left( {{\bf{G}}_k^{}{{\bf{W}}_l}} \right)}  + tr\left( {{\bf{G}}_k^{}{\bf{Q}}} \right) + {P_{ER,k}} + \sigma _{ER,k}^2 \le \\
&\hspace{3.8cm}{2^{{\bar{c}}_{k}}}(\ln(2){c_{k}} - \ln(2){\bar{c}}_{k} + 1).
\end{split}
\end{equation}
By replacing \eqref{SDRR_C3} and \eqref{SDRR_C4} with \eqref{SDRR_C3_Appro} and \eqref{SDRR_C4_Appro}, we obtain the convex approximation of problem \eqref{SDR_relax} at $\{{{\bar{b}}_j}, {{\bar{c}}_{k}}\}$ as follows
\begin{equation}
\label{SDR_appro}
\begin{split}
&\mathop {\max }\limits_{{{\bf{W}}_j},{\bf{Q}},r_j,a_j,b_j,c_{k},d_{k,j}, \forall k \in \mathcal{K}, \forall j \in \mathcal{J}} \sum\limits_{j\in \mathcal{J}}{\ln (r_j)} \\
&{\text{s.t.}} \, \eqref{SDRR_C1}, \eqref{SDRR_C2}, \eqref{SDRR_C5},\\
&\hspace{0.5cm} \, \eqref{SDRR_C3_Appro},\eqref{SDRR_C4_Appro},\\
&\hspace{0.5cm}\,  \eqref{SDR_C1}-\eqref{SDR_C4}.
\end{split}
\end{equation}
Then we can handle problem \eqref{SDR_relax} by using SCA approach in which the approximation problem \eqref{SDR_appro} is solved iteratively. More specifically, in the $(\kappa+1)$-th iteration, the following convex optimization problem is to be solved
\begin{subequations}
\label{SDR_SCA}
\begin{align}
&\mathop {\max }\limits_{{{\bf{W}}_j},{\bf{Q}},r_j,a_j,b_j,c_{k},d_{k,j}, \forall k \in \mathcal{K}, \forall j \in \mathcal{J} } \sum\limits_{j\in \mathcal{J}}{\ln (r_j)} \nonumber\\
&{\text{s.t.}} \, \eqref{SDRR_C1}, \eqref{SDRR_C2}, \eqref{SDRR_C5}, \nonumber\\
&\hspace{0.5cm} \, \sum\limits_{l \in \mathcal{J}, l \ne j} {tr\left( {{\bf{H}}_j^{}{{\bf{W}}_l}} \right)}  + tr\left( {{\bf{H}}_j^{}{\bf{Q}}} \right) + {P_{IR,j}} + \sigma _{IR,j}^2  \nonumber\\
&\hspace{2.5cm}\le  {2^{b^{(\kappa)}_j}}(\ln(2){b_j} - \ln(2){b^{(\kappa)}_j} + 1), \, \forall j \in \mathcal{J}, \label{SDR_SCA_C1} \\
&\hspace{0.5cm} \, \sum\limits_{l \in \mathcal{J}} {tr\left( {{\bf{G}}_k^{}{{\bf{W}}_l}} \right)}  + tr\left( {{\bf{G}}_k^{}{\bf{Q}}} \right) + {P_{ER,k}} + \sigma _{ER,k}^2  \nonumber\\
&\hspace{2.5cm} \le {2^{{c^{(\kappa)}_{k}}}}(\ln(2){c_{k}} - \ln(2){c^{(\kappa)}_{k}} + 1), \, \forall k \in \mathcal{K}, \label{SDR_SCA_C2}\\
&\hspace{0.5cm}\,  \eqref{SDR_C1}-\eqref{SDR_C4}. \nonumber
\end{align}
\end{subequations}
Once problem \eqref{SDR_SCA} is solved, the optimal solution to it can be used to construct the optimization problem in the next iteration.

For the SCA iterative process, initialization is necessary to be concerned. However, the non-convexity of the feasible set of problem \eqref{SDR_relax} causes difficulty in finding an initial feasible point in \eqref{SDR_relax} directly. Consequently, we propose the following convex feasibility problem
\begin{subequations}
\label{Feas}
\begin{align}
&\mathop {\min}\limits_{{{\bf{W}}_j},\forall j \in \mathcal{J},{\bf{Q}}} \, 0 \nonumber\\
&s.t. \,    {tr({\bf{H}}_j^{}{{\bf{W}}_j})}/\Gamma_{IR,j} \ge  \sum\limits_{l\in \mathcal{J}, l\ne j} {tr\left( {{\bf{H}}_j^{}{{\bf{W}}_l}} \right)}  + tr\left( {{\bf{H}}_j^{}{\bf{Q}}} \right) +   \nonumber\\
& \hspace{0.5cm} {P_{IR,j}} + \sigma _{IR,j}^2 , \, \forall j \in \mathcal{J}, \label{FeasC1} \\
&\hspace{0.5cm} tr({{\bf{G}}_k{{\bf{W}}_j}})=0, \forall k \in \mathcal{K}, \forall j \in \mathcal{J}\label{FeasC2}\\
&\hspace{0.5cm} \eqref{SDR_C1}-\eqref{SDR_C4}, \nonumber
\end{align}
\end{subequations}
where $\Gamma_{IR,j}$ is a small positive threshold for $\overline{SINR}_{IR,j}$. Let $\{{{{\bf{W}}_j^{(0)}},\forall j \in \mathcal{J},{\bf{Q}}^{(0)}}\}$ be a feasible point obtained by solving \eqref{Feas}. It can be observed that $\{{{{\bf{W}}_j^{(0)}},\forall j \in \mathcal{J},{\bf{Q}}^{(0)}}\}$ is also feasible in $\eqref{SDR}$. Because of the equivalent relationship between \eqref{SDR} and \eqref{SDR_relax}, the initial values of $b_j$ and $c_k$ for the SCA iterative process can be given by
\begin{subequations}
\label{bczero}
\begin{align}
&{b^{(0)}_j} = \log_2 {\bigg{(}} \sum\limits_{l \in \mathcal{J}, l \ne j} {tr\left( {{\bf{H}}_j^{}{{\hat{\bf{W}}}^{(0)}_l}} \right)}  + tr\left( {{\bf{H}}_j^{}{\hat{\bf{Q}}}^{(0)}} \right) + {P_{IR,j}} \nonumber\\
&\hspace{6.1cm}+ \sigma _{IR,j}^2 {\bigg{)}} , \\
&{c^{(0)}_{k}} = \log_2 {\bigg{(}} \sum\limits_{l \in \mathcal{J}} {tr\left( {{\bf{G}}_k^{}{{\hat{\bf{W}}}^{(0)}_l}} \right)}  + tr\left( {{\bf{G}}_k^{}{\hat{\bf{Q}}}^{(0)}} \right) + {P_{ER,k}} \nonumber\\
&\hspace{6cm}+ \sigma _{ER,k}^2 {\bigg{)}}.
\end{align}
\end{subequations}

According to the discussion above, we propose the following algorithm.
\begin{algorithm}[htb] \label{AL1}
\caption{Algorithm for solving problem \eqref{SDR}}

1. Initialize $\kappa:= 0$.

2. Obtain $\{{{{\bf{W}}_j^{(0)}},\forall j \in \mathcal{J},{\bf{Q}}^{(0)}}\}$ by solving the convex feasibility problem \eqref{Feas}.

3. Calculate ${b^{(0)}_j}, \forall j \in \mathcal{J}$ and ${c^{(0)}_k}, \forall k \in \mathcal{K}$ through \eqref{bczero}.

4. \textbf{Repeat}

5. \hspace{0.5cm} Solve \eqref{SDR_SCA} and obtain the optimal solution $\{{{\hat{\bf{W}}}^{(\kappa+1)}_j},{\hat{\bf{Q}}}^{(\kappa+1)},{\hat{r}}_j^{(\kappa+1)},{\hat{a}}_j^{(\kappa+1)},{\hat{b}}_j^{(\kappa+1)},{\hat{c}}_{k}^{(\kappa+1)},{\hat{d}}_{k,j}^{(\kappa+1)}\}$.

6. \hspace{0.5cm} Set ${b^{(\kappa+1)}_j}:={\hat{b}}_j^{(\kappa+1)}$ and ${c^{(\kappa+1)}_k}:={\hat{c}}_{k}^{(\kappa+1)}$.

7. \hspace{0.5cm} Set $\kappa:= \kappa + 1$.

8. \textbf{Until} convergence of the objective in \eqref{SDR_SCA}.


\end{algorithm}


If the solution $\{{{{\bf{W}}_j^{*}},\forall j \in \mathcal{J},{\bf{Q}}^{*}}\}$ obtained through Algorithm 1 satisfies $\rank({{\bf{W}}_j^{*}})=1$,
then the (sub)optimal information beamforming vectors for the original design problem can be obtained via employing EVD. Otherwise, the Gaussian randomization technique \cite{Ren19} can be applied to generate the solutions for information beamforming vectors. It is interesting to mention that we obtain rank-one ${{\bf{W}}_j^{*}}$ with high probability in our simulation trials. On the other hand, the rank of ${{\bf{Q}}_j^*}$ is always greater than 1.

\section{Simulation Results}
\label{sect:numerical}

In this section, we provide numerical results to illustrate the performance of the proposed beamforming algorithms. We assume that the path loss model is $P_L= (d/d_0)^\alpha $ for all users (FUs and MUs), where $d$ denotes
the distance between one given user to its connecting BS, $d_0$ is the
reference distance set to be $1$ m, and $\alpha = 3.5$ is the path
loss exponent. Moreover, we set $d_e = 6$m, $d_j = 12$m and $d_i = 30$m as the distances from FBS to ER FUs, IR FUs and MUs respectively.
All channel coefficients are modeled as i.i.d Rayleigh fading, i.e., the channel vectors ${\bf{h}}_j$, ${\bf{g}}_k$ and ${\bf{i}}_n$ follow the distribution $\mathcal{CN}({\bf{0}}, P_L{\bf{I}})$. As for the noise power, we set $\sigma^2_{IR,j} =  \sigma^2_{ER,k} =\sigma^2 ,\forall j \in \mathcal{J}, \forall k \in \mathcal{K}$. In addition, we assume $\varpi_k = \varpi, \forall k \in \mathcal{K}$ and $\eta_n = \eta, \forall n \in \mathcal{N}$ for the energy harvesting and interference threshold. The parameter value setting is provided in Table I.

\begin{table}
\centering
\caption{Simulation Parameters Setting}\label{tab:table}
\begin{tabular}{|p{6cm}|p{2cm}<{\centering}|}
\hline
\hline
Parameters  & Values   \\
\hline
Number of MUs $N$  & $2$  \\
\hline
Number of IR FUs $J$  & $2$  \\
\hline
Number of ER FUs $K$  & $2$  \\
\hline
Number of FBS's antennas $T$ & $6$ \\
\hline
While noise power $\sigma^2$ & $-50$dBm  \\
\hline
ER FUs' energy harvesting efficiency $\xi$ & $0.5$   \\
\hline
MUs' interference threshold $\eta$ & $100\sigma^2$ \\
\hline
MBSs' cross-tire interference to IR FUs $P_{IR,j}$ &  $20\sigma^2$   \\
\hline
MBSs' cross-tire interference to ER FUs $P_{ER,k}$ &  $20\sigma^2$   \\
\hline
\hline
\end{tabular}
\end{table}

\begin{figure} \centering
\label{fig:Converge}
\includegraphics[width=0.4\textwidth]{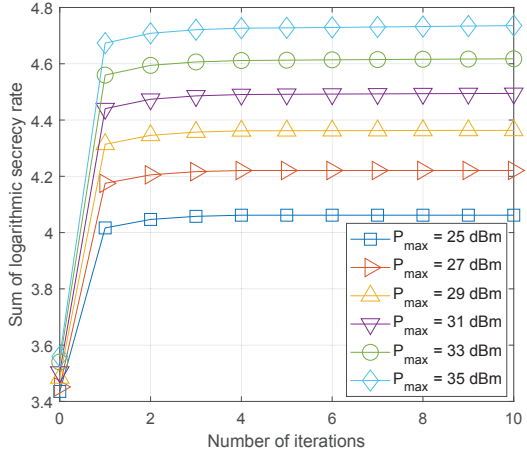}
\caption{Convergence of Algorithm 1 with different maximal transmit power for $\varpi = 1$mW.
}
\end{figure}

In Figure 1, we study the convergence of Algorithm 1 with different values of the maximal transmit power $P_{max}$ by setting ER FUs' energy harvesting threshold $\varpi$ as $1$mW. As observed from Figure 1, no matter how $P_{max}$ is changed, the sum logarithmic secrecy rates increases
monotonically before convergence and Algorithm 1 can converge in six iterations. 

\begin{figure} \centering
\label{fig:SCthroughput:powerup}
\includegraphics[width=0.4\textwidth]{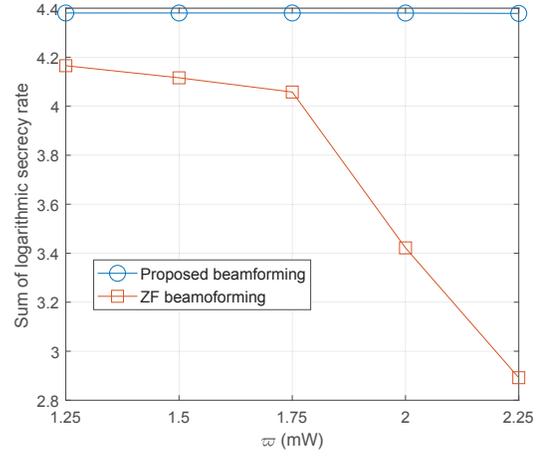}
\caption{Sum of logarithmic secrecy rates achieved by the proposed beamforming and ZF beamforming designs versus $\varpi$ with $P_{max} = 30$dBm.}
\end{figure}

Figure 2 demonstrates the sum of logarithmic secrecy rates achieved by the proposed beamforming design as a function of  $\varpi$ with $P_{max} = 30$dBm. For the purpose of comparison, we also examine the performance of the beamforming design which is based on the zero-forcing (ZF) approach. In the ZF scheme, the direction of the information beamforming vector for the $j$-th IR FU is predetermined as $\left({\bf{I}}- {{\bf{H}}^\dagger_{ZF,j}{{\bf{H}}_{ZF,j}}}\right) {\bf{h}}_j$ in order to eliminate the interferences to the other IR FUs and the information leakage to all ER FUs, where ${{\bf{H}}_{ZF,j}} = [{{\bf{h}}_1},...,{{\bf{h}}_{j - 1}},{{\bf{h}}_{j + 1}}$ $,...,{{\bf{h}}_J},{{\bf{g}}_1},...,{{\bf{g}}_K}]^H$. On the other hand, the energy covariance matrix is designed such that it causes no interferences to IR FUs, i.e., ${\bf{h}}_j^H{\bf{Q}}{\bf{h}}_j=0, \forall j \in \mathcal{J}$. With the ZF scheme, the original design problem \eqref{mainpro} can be simplified into a convex problem where the powers of the information beamforming vectors and the energy covariance matrix are optimized. We can observe that the proposed beamforming design outperforms the ZF beamforming design. In addition, when $\varpi$ is greater than 1.75 mW, the sum of logarithmic secrecy rates achieved by the ZF beamforming design degrades significantly. By contrast, the performance of the proposed beamforming design degrades slightly, which reveals the effectiveness of the proposed beamforming design.

\section{Conclusions}
\label{sect:conclusion}
In this paper, we considered a MISO Femtocell which is overlaid with a Macrocell in co-channel deployment. A sum logarithmic secrecy rate maximization beamforming design problem with interference constraints for MUs and EH constraints for
ER FUs was formulated. To deal with the beamfomring design problem, we proposed an algorithm based on the SDR and SCA techniques. The
convergence of the proposed algorithm and the effectiveness of
the proposed beamforming design were illustrated in simulation
results.


\ifCLASSOPTIONcaptionsoff
  \newpage
\fi

\bibliographystyle{IEEEtran}
\bibliography{IEEEabrv,refabrv}

\end{document}